\documentclass[11pt, a4paper,twocolumn]{article}

\usepackage{graphicx}
\usepackage{amsmath}
\usepackage{amsfonts}
\usepackage{amssymb}
\usepackage{hyperref}
\usepackage{slashed}

\newcommand{\be}{\begin{equation}}
\newcommand{\ee}{\end{equation}}
\newcommand{\bea}{\begin{eqnarray}}
\newcommand{\eea}{\end{eqnarray}}
\newcommand{\ba}{\begin{array}{ll}}
\newcommand{\ea}{\end{array}}

\begin{document}

\twocolumn[
  \begin{@twocolumnfalse}
\def\lsim{\mathrel{\raise.3ex\hbox{$<$\kern-.75em\lower1ex\hbox{$\sim$}}}}
\def\gsim{\mathrel{\raise.3ex\hbox{$>$\kern-.75em\lower1ex\hbox{$\sim$}}}}
\sf
\centerline{\LARGE Supersymmetry and Unification:}
\centerline{\LARGE Heavy Top Was the Key$^{\S}$ }
\vspace{5mm}
\centerline{\large Goran Senjanovi\' c}
\centerline{{\it ICTP, Trieste, Italy }}
\vspace{5mm}
\centerline{\large\sc Abstract}
\begin{quote}
\small
I review the unification of gauge couplings of strong, weak and electro-magnetic interactions. I start by recalling the history of the most
important prediction of low-energy supersymmetry: the correct value of the weak mixing angle tied to 
a large top quark mass. I then turn to the
discussion of the present day situation of the minimal supersymmetric Grand
Unified Theories based on $SU(5)$ and $SO(10)$ groups, and I show why the minimal $SU(5)$ is in accord with experiment. For the sake of completeness I also summarize the problems
and possible solutions of the minimal ordinary $SU(5)$. One version, based on the minimal Georgi-Glashow model, augmented by
the adjoint fermion, predicts a light fermion triplet to lie below TeV or so. Its (lepton number violating) decays offer
a hope of probing neutrino (Majorana) masses and mixings at the LHC. 
\end{quote}
  \end{@twocolumnfalse}]
 {
 \renewcommand{\thefootnote}%
   {\fnsymbol{footnote}}
 \footnotetext[4]{To be published in the proceedings of the Scientific and Human Legacy of Julius Wess,
 Memorial Workshop held in Donji Milanovac, Serbia, August 2011.  
 There is plagiarism in my title. It
is borrowed from the beautiful historical overview of the V-A theory
by Weinberg~\cite{Weinberg:2009zz}.

}
}

 \rm
\newpage
\tableofcontents

\section{\large Foreword}

There can be no pleasure contributing to a memorial volume,
especially if it is dedicated to a great physicist and an
outstanding human being. It is a great honor though. I got to know
Julius Wess some thirty years ago, and he was a good friend. He was also
a great friend of Balkan physics, especially what once upon a time
was called Yugoslav physics. The passage of time naturally brings
nostalgia, and as I am writing this I cannot help but feel nostalgia
for the golden days of our spring when we met for the first time. We
both lectured in the 1983 Adriatic School in Dubrovnik, an excellent
event organized jointly by the Belgrade and Zagreb High Energy
groups. Julius was immediately hooked to the beauty of Dubrovnik,
and fell in love with the atmosphere of hospitality and friendliness
which he associated with the old Yugoslavia. He kept coming back
regularly to both Croatia and Serbia and it is more than appropriate
that his memorial workshop was held in the Djerdap National Park,
one of the jewels of Serbia.

Julius Wess, as everybody knows, was one of the fathers of supersymmetry,
which for the last four decades has been shaping our field, for
better or for worse. I was introduced to supersymmetry in 1974, just
as I was starting my Ph. D. research, when I heard a talk by Bruno
Zumino at Columbia University. He and Julius had just finished their
classic work, and Zumino gave an inspiring talk. I for one was
enchanted by the beauty of supersymmetry, although I could share
Zumino's sentiment when he compared supersymmetry to a giraffe,
saying: "I am sure we all agree that a giraffe is truly beautiful,
but she doesn't seem to serve any purpose".  As much as he was right
about everything else in his talk, in this he was wrong: giraffes,
as all  living creatures, have an important role in preserving the
ecological balance of nature. It is ironic that over the years
supersymmetry would become the main candidate for the theory
beyond the Standard Model. If anything, it found probably too much
of a practical role, and today is not even considered "exotics" at
the LHC. But then, it is neither first nor the last time that heresy becomes dogma.

\section{\large Unification of gauge couplings: low energy supersymmetry and heavy top}

Around the same time, in 1974, Grand Unification started to emerge
as a natural next step of the Standard Model unification, and
brought in its huge scale of at least $10^{15}$ GeV as dictated by
proton stability. This made the so-called hierarchy question of the
smallness of the electroweak scale compared to the Planck scale much
more acute, for Grand Unification also brought in  extreme
fine-tuning, at least in its minimal versions. In the next five
years or so, the Standard Model was to emerge triumphantly  as the
theory of both the charged and neutral weak processes. The theory agreed with all the data for
the value of the weak mixing angle $\sin^2\theta_W(M_W) \simeq
0.21$. This fit perfectly with the minimal $SU(5)$ theory of Georgi and
Glashow~\cite{Georgi:1974sy}, after Georgi, Quinn and
Weinberg\cite{Georgi:1974yf} taught us how to compute
$\sin^2\theta_W(M_W)$. They also taught us how to compute the unification
scale and thus the proton lifetime, around $10^{30}\,yr$ or so. As
Maurice Goldhaber put it nicely, many experimentalists then rushed
underground all across the globe, to look for proton decay.

All looked well, not a cloud on the horizon, except for the growing
uneasiness about the hierarchy problem. And then it became clear
that low energy supersymmetry around TeV or so (notice {\it or so},
we will come back to this later) would, if not cure the hierarchy
problem, alleviate it greatly for you could do your fine-tuning at
the tree level and live happily ever after. 
 
 However, it was soon pointed out
 by Dimopoulos, Raby and Wilczeck\cite{drw} (and, we will
learn later, independently by Ib\'a\~nez and Ross\cite{Ibanez:yh}) that
that the scale
of unification and predicted proton lifetime get significantly increased 
 in the minimal supersymmetric GUTs, while
$\sin^2\theta_W(M_W)$ remains almost unaffected.

Inspired by Dimopoulos et.al., in the fall of 1981, Bill
Marciano and I decided to do a careful two-loop analysis of the
unification with low energy supersymmetry,\cite{Marciano:1981un} and
found out that Einhorn and Jones were doing it already (their
paper\cite{Einhorn:1981sx} appropriately appeared before ours). The
result was an increased $\sin^2\theta_W (M_W ) \simeq 0.23$, in clear
disagreement with claimed value $\sin^2\theta_W (M_W )_{\rm exp}
\simeq 0.21$. Bill and I liked the idea of low energy supersymmetry
and decided to stick our necks out in its favor. First of all,
 if you took into account the $\rho$
parameter of the Standard Model, the fit was actually
 \begin{equation}
 \sin^2\theta_W (M_W ) \simeq 0.22 - 0.49 (1-\rho^2).
 \end{equation}
In the Standard Model, with radiative corrections included, the
(wrongly)  quoted value  was $\rho \simeq 0.98$, assuming that the
top quark was going to be light according to the prejudice of the
day. However, with a heavy top quark, around $200$ GeV or so, you
would get  $\rho \simeq 1.01$, and $\sin^2\theta_W (M_W )_{\rm exp}
\simeq 0.23$. Bill and I then argued in favor of low energy supersymmetry and a 
heavy top, and thus predicted the correct value of
$\sin^2\theta_W (M_W )$ ten years before LEP was to confirm it.
Moreover, the top, lo and behold, turned out to be heavy. Looking
back, I feel both proud and happy that we had the courage and sense
to say it in print.

{\it Historia magistra vitae}, says the Latin proverb: history is
the teacher of life. I don't know what this little history teaches
us, but I think it was appropriate to bring it up. After all, it is
not well known, not enough, that the successful unification, the
greatest success of low energy supersymmetry, went hand in hand with
the heavy top quark, long before experiment.

In what follows, I discuss the implementation of gauge coupling
unification with grand unified theories. I argue that the TeV scale
of supersymmetry ought be taken as soft, and that there is no
guarantee that supersymmetry lies at the LHC reach. There was too
much optimism in arguing so, and a negative result should not be
taken against this beautiful idea, but rather against false
prophets.

\section{\large Minimal SU(5) }

It was the minimal SU(5) that caused the underground rush, for it
predicted proton lifetime on the order of $\tau_p \simeq 10^{30}$
yr. And on top, it also predicted the nucleon decay branching ratios
as shown by Mohapatra~\cite{Mohapatra:1979yj}. Unfortunately these
predictions resulted from the wrong mass relations: $m_e = m_d$.
 These relations can be corrected
easily by simply adding higher dimensional operators\cite{Ellis:1979fg}, but then the theory stops being
predictive.

In any case this is only history now, for the theory is not 
consistent, since
\begin{itemize}
\item
gauge couplings do not unify
\item
neutrinos stay massless (as in the MSM).

\end{itemize}

Possible higher dimensional operators are not enough: neutrino mass
comes out too small. It is important to know
then what the minimal consistent realistic extensions are. Two such models were
constructed, by adding

   \begin{itemize}

   \item a symmetric complex scalar field~\cite{Dorsner:2005fq} (and higher dimensional operators for charged fermion masses) 

     $15_H = (1_C, 3_W) + (6_C, 1_W) + (3_C, 2_W)$,
     
      with $(1_C, 3_W)$ being the usual
     Higgs triplet behind the type II seesaw~\cite{Magg:1980ut}. 
The leptoquarks $(3_C, 2_W)$ may remain light (but not necessarily)
and offer an interesting phenomenology~\cite{Dorsner:2009mq}. There
is also an important prediction, a fast proton decay on the edge of
experimental limits.

       \item an adjoint fermion field\cite{Bajc:2006ia}

         $24_F=  (8_C, 1_W) + (1_C, 3_W) + (1_C, 1_W) + ...$

        The fields   $S=(1_C, 1_W)$ and $T=(1_C, 3_W)$ are responsible for type I\cite{seesaw} + III\cite{Foot:1988aq} hybrid seesaw.  
      Again, one needs higher dimensional operators both for charged femions and for realistic neutrino Dirac yukawa couplings

 This model predicts a light fermion triplet  T, with a mass below TeV so that the running of the SU(2)
 gauge coupling is slowed down and meets U(1) above $10^{15}$ GeV.  Its phenomenology is quite interesting for
 it leads to lepton number violation at colliders in the form of same sign di-leptons as suggested originally
 long ago~\cite{Keung:1983uu}. For the relevant studies in the context of the type III seesaw,
 see Ref.~\cite{Franceschini:2008pz}, and for this specific model, see Ref.~\cite{Arhrib:2009mz}.

\end{itemize}

   The two theories have in common a "fast" proton decay, with $\tau_p \leq 10^{35}$ yr.

\section{\large Minimal supersymmetric SU(5)}

 The underground rush continued, or better to say, got boosted with the success of the minimal supersymmetric
 SU(5).  As discussed in the Introduction, low energy supersymmetry, suggested in order to stabilize the Higgs mass hierarchy predicted correctly
 $\sin^2 \theta_W \simeq 0.23$,
  ten years before its confirmation at LEP.

    The unification scale too was predicted:  $M_{GUT} \simeq 10^{16} \mbox{ GeV}$ and in turn {$\tau_p (d=6) \simeq 10^{36 \pm 1}$ yr,
    which would have rendered proton decay out of reach of experiment.
However, supersymmetry leads to new contribution:  $d=5$
operators\cite{Sakai:1981pk} through the exchange of heavy color
triplet Higgsino ($T$ and $\bar T$). A rough estimate gives
\begin{equation}
   G_T   \simeq \frac{\alpha}{4 \pi} y_u \, y_d \frac{m_{\mbox{\tiny gaugino}}}{M_T m^2_{\tilde f}} \simeq 10^{-30} \mbox{
   GeV}^{-2},
\end{equation}
 which for
$y_u \simeq y_d \simeq 10^{-4}$, $m_{\mbox{\tiny gaugino}} \simeq
100$ GeV, $m_{\tilde f} \simeq $TeV and $M_T \simeq 10^{16} $ GeV
gives $\tau_p (d=5) \simeq 10^{30 - 31}$ yr.  It would seem that
today this theory is ruled out. It was actually proclaimed dead~\cite{Murayama:2001ur}, but then caution was
raised for two important reasons: i) the uncertainty in sfermion
masses and mixings\cite{Bajc:2002bv} and ii) uncertainty in
$M_T$\cite{Bajc:2002pg} due to necessity of higher dimensional
operators\cite{Ellis:1979fg} to correct bad fermion mass relations
$m_d = m_\ell$. The $d=4$ operators, besides correcting these
relations also split the masses of $m_3$ and $m_8$ of  the weak
triplet and color octet, respectively, in the adjoint $24_H$ Higgs
super multiplet. One gets
\begin{equation}
\left. \begin{array}{l}
M_{GUT} = M^0_{GUT} \left(  \frac{M_{GUT}^0}{\sqrt{m_3m_8}}\right)^{1/2}  \\
M_T = M_T^0 \left(\frac{m_3}{m_8} \right)^{5/2}
\end{array} \right\} M^0_{GUT} \simeq 10^{16} \mbox{ GeV},
\end{equation}
where $M^0_{GUT}$ denotes the GUT scale predicted for $m_3=m_8$ at
the tree level with d=5 operators neglected. The fact that $M_{GUT}$
goes up by lowering $m_{3,8}$  was noticed quite some time
ago~\cite{Bachas:1995yt}. In principle, the ratio of the triplet and
octet masses can be as large as one wishes, so at first glance the
proton lifetime would seem not to be limited from above at all.
However, in recent years, Dvali\cite{Dvali:2007hz} has been arguing
that the effective strong gravity scale is given by
$\Lambda_{\text{gr}} = M_{Pl}/\sqrt{N}$, where $M_{Pl} \simeq
10^{18}$ and $N$ is the number of states for a theory in question.
With $N$ around 100, in order for this theory to be perturbative,
one must demand $M_{\text{GUT}} \lesssim 10^{17} \text{ GeV}$ or
$\tau_p \lesssim 10^{35} \text{ yrs}$.

  In short, the minimal supersymmetric $SU(5)$ is still a perfectly viable theory, and the $d=5$ proton decay is close
  to the present limit.
  This minimal theory must account for neutrino masses and mixings and
  the only possibility is through the violation of R-parity.
       The first important implication of not assuming R-parity  is that the lightest neutralino cannot be dark matter  for it
     decays too fast with the collider signature of lepton number violation. Thus the only dark matter candidate is
     an unstable gravitino~\cite{Takayama:2000uz},
and it works perfectly in the MSSM, see~\cite{Bajc:2010qj}.

\section{\large SO(10)}

Although $SU(5)$ is the minimal theory of grand unification and as such deserves maximal attention as
a laboratory for studying  proton decay, $SO(10)$ has important merits

\begin{itemize}

 \item it unifies a fermion family in a spinorial $16_F$ representation and as such is a minimal unified theory
of matter and interactions,
   \item it automatically contains right-handed neutrinos   $N$,

   \item it gives naturally  $ M_N \gg M_W$ and so neutrino has a tiny mass through the see-saw mechanism,

 \item in supersymmetry R-parity is a gauge symmetry~\cite{rparity},

 \item in the renormalizable version R-parity
  remains exact~\cite{Aulakh:2000sn} (as in any
 B-L symmetric theory~\cite{Aulakh:1999cd}), implying the stable
  lightest superpartner, natural
 dark matter candidate.

\end{itemize}

All the fermion
masses and mixings can be accounted for with the $10_H$ and $126_H$ representations, the latter providing
a large mass to the right-handed neutrinos.
While the non supersymmetric, SO(10) theory was studied at length over the years, no predictive realistic model
of fermion masses and mixings ever emerged.
 
 If $10_H$ is taken to be real, the predictions turn out to be
wrong~\cite{Bajc:2005zf}, and making it complex doubles the Yukawas and kills the
predictions.

This situation improves in supersymmetry, since
 holomorphicity guarantees single Yukawas for both $10_H$
and $126_H$. The minimal supersymmetric version,
coined renormalizable,  suggested already in
1982~\cite{Senjanovic:2009kr}and rediscussed ten years
later~\cite{Babu:1992ia}, has been under scrutiny in recent
years~\cite{susyso10}. An important  boost was provided by
an observation that $b - \tau$ unification, tied
with the type II seesaw, gives naturally large atmospheric mixing angle~\cite{Bajc:2002iw}.
Moreover, it predicts $\theta_{13} \simeq 10^\circ$~\cite{Goh:2003sy}, in accord with T2K~\cite{Abe:2011sj} and Daya Bay~\cite{An:2012eh}
any need for flavour symmetries

After a great initial success, when pinned down precisely, the
theory ran into tension between fast proton and neutrino masses. It
was revisited~\cite{Bajc:2008dc} and shown to work with the so-called
split supersymmetry spectrum of heavy
sfermions~\cite{ArkaniHamed:2004fb}. Although by splitting the
supersymmetric spectrum one loses the hierarchy protection, the
unification of gauge couplings is preserved.
 If one accepts split supersymmetry, one
can also develop a realistic model~\cite{Bajc:2004hr} based on a beautiful
radiative mechanism~\cite{Witten:1979nr}.

Instead of a large $126_H$ Higgs, one may choose a $16_H$ and then build $126_H$ Higgs effectively as
$16_H^2$. This induces a proliferation of couplings and needs extra flavor symmetries, and thus goes
beyond a simple GUT picture; see e.g.\cite{Babu:1998wi}.

\section{\large Conclusions}

I gave here an account of the unification of gauge couplings in supersymmetry, and showed how it
prophetically predicted a heavy top quark, some ten years before LEP and some fifteen years before the top
discovery. This still remains the main prediction of low energy supersymmetry, giving us hope to discover it at the LHC.
We should keep in mind though that it can easily be above the LHC reach, without losing unification, only at the expense
of some additional fine-tuning. I, for one, can live with that.

I then discussed the minimal supersymmetric grand unified theories
based on SU(5) and SO(10) gauge groups as a way of completing the
program of unification. It is remarkable that the minimal
supersymmetric SU(5) is still in accord with experiment. I showed,
on the other hand, that the computation of the proton decay
branching ratios can be achieved in the SO(10) theory with split
supersymmetry. For the lovers of unification who do not worry about
the hierarchy problem this is OK, for splitting the supersymmetry
preserves the unification of gauge couplings. I was rather
telegraphic here for the lack of space; a reader who wants to learn
more about this beautiful subject can have a look at my recent
comprehensive review~\cite{Senjanovic:2011zz}.


\section{\large Outlook}

We all know how quickly a heresy can turn into a dogma, and I must
admit that the more supersymmetry was becoming a dogma over the
years, the less I felt my original enthusiasm. Moreover, it was and still is, hard not to have 
doubts in a theory whose main prediction went hand in hand with a 
a desert in energies all the way to the unreachable GUT scale. Now that,
according to many, it is getting into trouble I start to feel my
old love again, and wish to defend it. It is more than
an appropriate to paraphrase Mark Twain here:  any rumor
about the death of low energy supersymmetry has been greatly
exaggerated.

I put my money where my mouth is and in 2003, at a conference
held in Vrnja\v{c}ka Banja, Serbia, I made a bet with
Julius that supersymmetry would be discovered at the LHC. He bet
against it, so that either way it was a win-win situation for him
because he was going to gain much more by
losing the bet (btw, it was a serious bet, a
case of expensive wine). It may appear strange that I would bet in
favor of supersymmetry in view of my doubts. But then again, I too have a vested
interest in low energy supersymmetry, and either outcome would be a
win for me as well. I only regret deeply that Julius will not be
around to settle the bet.

Julius Wess will not only be missed by family, friends, colleagues
and collaborators, but by many of us from the old Yugoslavia who
recall with warmth his love and support for these countries. I for
one will miss very much seeing him, as happened many times, in these
lands so dear to me.

\section*{\large Acknowledgments}

I am deeply grateful to the organizers of the Julius Wess Memorial
Workshop for giving me the opportunity to honor a great physicist.
It would take too much space to thank all of my collaborators on the
topics discussed here, to whom I am deeply indebted.  I wish though
to acknowledge with great pleasure my original collaboration with
Bill Marciano on the issue of supersymmetric unification. I thank Bill and Borut Bajc
for careful reading of this manuscript and for useful comments. Special
thanks are due to Alejandra Melfo and Miha Nemev\v{s}ek for
discussions and help with this write-up.

\end{document}